\documentstyle[multicol,aps,epsf]{revtex}

\begin{document}
\title{Survival Probability in a Random Velocity Field}
\author{S.~Redner}

\address{Center for Polymer Studies and Department of Physics, Boston
University, Boston, MA, 02215}
\maketitle
\begin{abstract}
 
The time dependence of the survival probability, $S(t)$, is determined
for diffusing particles in two dimensions which are also driven by a
random unidirectional zero-mean velocity field, $v_x(y)$.  For a
semi-infinite system with unbounded $y$ and $x>0$, and with particle
absorption at $x=0$, a qualitative argument is presented which indicates
that $S(t)\sim t^{-1/4}$.  This prediction is supported by numerical
simulations.  A heuristic argument is also given which suggests that the
longitudinal probability distribution of the surviving particles has the
scaling form $P(x,t)\sim t^{-1}u^{1/3}g(u)$.  Here the scaling variable
$u\propto x/t^{3/4}$, so that the overall time dependence of $P(x,t)$ is
proportional to $t^{-5/4}$, and the scaling function $g(u)$ has the
limiting dependences $g(u)\propto{\rm const.}$ as $u\to 0$ and
$g(u)\sim\exp(-u^{4/3})$ as $u\to\infty$.  This argument also suggests
an effective continuum equation of motion for the infinite system which
reproduces the correct asymptotic longitudinal probability distribution.

\end{abstract}
\begin{multicols}{2}

\section{INTRODUCTION}

Consider a diffusing particle in the semi-infinite planar domain
$(x>0,y)$ which is absorbed when $x=0$ is reached.  In addition to the
diffusion, the particle is driven by a unidirectional random velocity
field in which $v_x(y)$ is a random, zero-mean function of $y$ only
(Fig.~1).  This type of stochastic motion was introduced by Matheron and
de Marsily (MdM)\cite{mm} to describe the hydrodynamic dispersion of
dynamically-neutral tracer in a sedimentary layered rock formation.
Although the longitudinal bias averaged over an infinite number of
transverse layers is zero, the typical bias over a finite number of
layers is a fluctuating quantity which is a decreasing function of the
number of layers that a random walk visits.  This non-vanishing residual
bias underlies the faster than diffusive transport of the model.  In an
infinite system it has been established that the typical horizontal
displacement $x_{\rm typ}\propto t^{3/4}$\cite{bou,sr,zum} and that the
configuration averaged distribution of longitudinal displacements has
the form $P(x,t)\sim t^{-3/4}\,\exp[-(x/t^{3/4})^{4/3}]$.  There are
also strong fluctuations in the probability distribution between
different samples of the velocity field, as well as a very slow
convergence to the asymptotic limit.

While much is now understood about transport in the MdM model, we wish
to investigate its first passage properties.  Specifically, we consider
the semi-infinite two-dimensional system $x>0$ with an absorbing
boundary at $x=0$, and study the time dependence of the particle
survival probability, $S(t)$.  The survival probability in a finite
system with absorbing boundaries at $x=\pm L$ and the same
unidirectional random velocity field has been studied
previously\cite{sr}; however this system exhibits fundamentally
different behavior than the semi-infinite system that is treated here.

In the absence of a velocity field, it is well known that in the
semi-infinite system $S(t)$ asymptotically decays in time as
$t^{-1/2}$\cite{rw}.  Because the velocity field in the MdM model has no
net longitudinal bias, it is not immediately obvious how the behavior of
$S(t)$ will be affected.  Naively, one might expect that the dominant
contribution to $S(t)$ will arise from those velocity configurations
whose average bias is directed away from the boundary.  This is indeed
the case, and from this starting point, we present a simple argument
which suggests that the survival probability, averaged over all
realization of the velocity field, is proportional to $t^{-1/4}$.  This
prediction is in excellent agreement with our numerical results.

\begin{figure}
\narrowtext
\epsfxsize=2.5in\epsfysize=1.95in
\hskip 0.3in\epsfbox{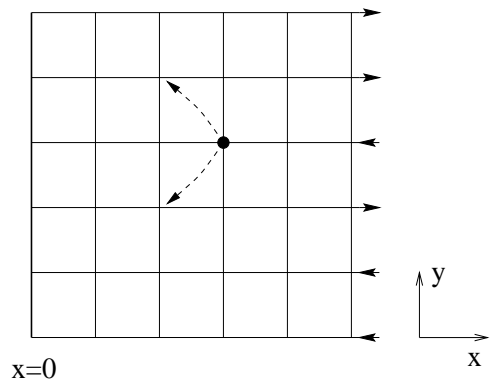}
\vskip 0.15in
\caption{The random velocity field in a realization of the  MdM model on
a finite width square lattice strip.  On the horizontal bonds, the
direction of the velocity field is indicated by the arrows.  In a single
time step, a particle (heavy dot) can move equiprobably only to one of
the two target sites indicated.  Particles are absorbed at $x=0$.
\label{fig1}}
\end{figure}

Interestingly, a $t^{-1/4}$ decay of the survival probability has been
found previously for diffusing particles in a semi-infinite two
dimensional system with a unidirectional zero-bias, but deterministic
velocity field of the form $v_x(y)=-v_x(-y)$\cite{wind}.  Although the
mechanism that leads to $S(t)\sim t^{-1/4}$ in this class of velocity
fields is different than that for the MdM model, the two systems share
the feature that their velocity fields have no average bias in $x$.  It
would be interesting to determine whether a $t^{-1/4}$ decay of the
survival probability is characteristic of all semi-infinite systems with
a zero-mean longitudinal velocity field.

In the next section, we give a heuristic argument which suggests that
$S(t)\propto t^{-1/4}$.  We then present corroborating numerical
simulations in Sec.\ III.  We also find that the spatial probability
distribution of the surviving particles provides insight into the
formulation of a continuum equation of motion for the longitudinal
probability distribution in an unbounded geometry.  Thus in Sec.\ IV, we
infer this equation of motion and, from its solution, determine the
correct asymptotic longitudinal probability distribution in the
unbounded geometry.  We conclude with a brief discussion in Sec.\ V.

\section{CONFIGURATION-AVERAGED SURVIVAL PROBABILITY}

We first present our argument for the the time dependence of $S(t)$.
The basic idea is that the dominant contribution to this average arises
from the subset of all velocity configurations whose net bias is away
from the boundary, {\it i.\ e.}, in the $+x$-direction.  Conversely,
configurations with a bias along $-x$ will give individual contributions
to $S(t)$ which decay exponentially in time and thus should be
asymptotically negligible.

To determine which of the positively-biased velocity configurations give
the dominant contribution to $S(t)$, consider a discrete realization of
the MdM model on the square lattice in which the velocity is either
$+v_0$ or $-v_0$ with equal probability for a given value of $y$
(Fig.~1).  Periodic boundary conditions in the transverse direction are
employed, so that the system is a semi-infinite cylinder consisting of
$w$ rows.  For concreteness, the initial condition is
$p(x,y,t=0)={1\over w}\delta_{x,\ell}$, {\it i.\ e.}, a ring of
particles is initially placed at $x=\ell$, where $\ell$ is the lattice
spacing.  In this system, the probability that there are $n_+$
positively biased rows and $n_-$ negatively biased rows is ${\cal
P}(m)\propto {1\over\sqrt{w}}e^{-m^2/w}$, with $m=n_+-n_-$.  In a time
$t$, the number of layers visited by a random walk is
$w\propto\sqrt{Dt}/\ell$, where $D$ is the transverse (microscopic)
diffusivity.  By transforming from $m$ to the velocity $v=mv_0/w$, the
distribution of velocities within $w$ layers is
\begin{equation}
{\cal P}(v)\propto {1\over v_0}
\left({Dt\over\ell^2}\right)^{1/4}\exp\left[{-\left({v\over v_0}\right)^2
\left({Dt\over\ell^2}\right)^{1/2}}\right].
\label{pv}
\end{equation}
Because this distribution is strongly cut off when the argument of the
exponential is greater than one, we expect that the dominant
contribution to $S(t)$ will arise from those velocity configurations
whose net bias is within the range $0<v<v_0(\ell^2/Dt)^{1/4}$.

For a positively biased velocity configuration, we now estimate the
residual survival probability in the long time limit under the
assumption that $w$ is finite.  In this case, the particle will
uniformly sample the transverse extent of the system and it is sensible
to characterize the bias by its mean value $v$.  If $v$ is small, or
more properly the Peclet number $vx_0/D$ is small, then for $t<D/v^2$
the bias is irrelevant and consequently $S(t)\sim x_0/\sqrt{Dt}$, where
$x_0$ is the initial position of the particle\cite{rw}.  However, for
$t>D/v^2$, convection dominates and returns to the origin become
extremely unlikely.  Hence $S(t)$ should ``stick'' at the value attained
when $t=D/v^2$.  This implies that the asymptotic behavior of the
survival probability is simply, $S(t=\infty)\propto vx_0/D$.  This same
result can also be obtained with additional effort from a more rigorous
approach in which one solves the one-dimensional convection-diffusion
equation on the domain $x>0$ with the initial condition
$P(x,t=0)=\delta(x-x_0)$ and then computes $S(t)$ by integrating this
probability density over all $x>0$.

For the MdM model, we now average over all relevant velocity
configurations to determine $S(t)$.  That is
\begin{eqnarray}
S(t) &\sim &\int_0^\infty {vx_0\over D} {1\over
v_0} \left({Dt\over\ell^2}\right)^{1/4} \exp\left[{-\left({v\over v_0}\right)^2
\left({Dt\over\ell^2}\right)^{1/2}}\right]\, dv,\nonumber \\
& =&{1\over 2}{v_0x_0\over D} \left({\ell^2\over{Dt}}\right)^{1/4}\nonumber \\
& \propto& t^{-1/4}.
\label{pvav}
\end{eqnarray}

\section{NUMERICAL SIMULATIONS}

We now present numerical evidence to support the prediction that
$S(t)\propto t^{-1/4}$.  For simulating the particle
motion, we propagate the probability distribution for each velocity
configuration exactly\cite{havlin}.  The microscopic motion is defined
by the rule that in a single time step, a particle hops equiprobably by
$\pm\ell$ in the $y$-direction and by a distance sign$(v(y))\ell$ in the
$x$-direction on the square lattice, where $v(y)$ is the velocity field
at the $y$-coordinate of the particle before the hopping event (Fig.~1).
Accordingly, the probability that a particle is at $(x,y)$ at time $t$
evolves according to
\begin{eqnarray}
p(x,y,t+1)&={1\over 2}\,p(x-\ell\times{\rm sign}(v(y-\ell)),y-\ell,t)\nonumber \\
{} &  \,+ {1\over 2}\,p(x-\ell\times{\rm sign}(v(y+\ell)),y+\ell,t).
\label{evol}
\end{eqnarray}
For each velocity configuration, probability propagation yields the
exact distribution in the presence of the absorbing boundary
up to the maximum time specified.

We have performed the average over velocity configurations in two
complementary ways -- either an average over a representative sample of
velocity configurations, or an exhaustive average over all velocity
configurations for relatively small systems.  The former is
straightforward to implement, but the effect of rare configurations on
the results is unknown.  The latter, on the other hand, gives the exact
result for an infinite system, albeit only for short times.  This
exactness allows one to test for systematic trends in the data, an
analysis which is not feasible when averaging over a representative set
of velocity configurations.

A typical result for $S(t)$ up to $t=4095$, based on an average over 50
velocity configurations on a cylinder of width 400, is shown in Fig.~2.
The average bias of these 50 configurations turns out to be $-0.004$.
Beyond approximately 50 time steps, the data for the survival
probability is quite linear and a least-squares power law fit to the
data in this time range yields the exponent of $-0.2491$.  Further, the
slope between successive data points, or local exponent estimate
\begin{equation}
\alpha_w(t)\equiv
{\ln\left(S(t)/S(t-1)\right) \over
{\ln(t/(t-1))}},
\label{localexp}
\end{equation}
deviates from $-0.25$ by less than 0.002 for $t\agt 50$ (inset).  To
check that finite width effects do not substantially affect the results,
we also considered shorter times, $t\alt 1600$ and a slightly narrower
system ($w=300$), and averaged over 250 realizations.  The average bias
of these configurations turns out to be $+0.002$.  This case yielded a
best fit exponent of $-0.2503$.  These two data sets strongly suggest
that $S(t)\propto t^{-1/4}$ in the long time limit.  However, because
the average is performed only over an infinitesimal fraction of all
velocity configurations, it is possible that extreme configurations
could alter the results.  For this reason, we now investigate the exact
behavior of $S(t)$ for short times by averaging over all velocity
configurations.

\begin{figure}
\narrowtext
\epsfxsize=3.0in\epsfysize=3.0in
\hskip 0.05in\epsfbox{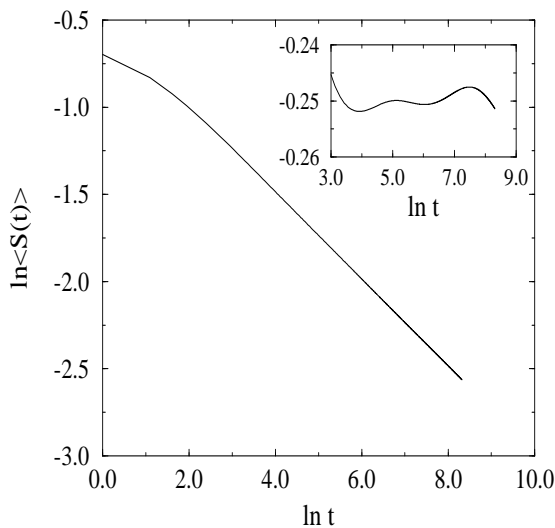}
\vskip 0.1in
\caption{Plot of $S(t)$ versus $t$ on a double
logarithmic scale, based on averaging the exact probability distribution
over a finite number of velocity configurations for the system discussed
in the text.  The inset shows the local slopes between neighboring data
points. 
\label{fig2}}
\end{figure}

In this complete enumeration, we consider odd values of $w$ for
convenience.  Because of the periodic transverse boundary conditions,
many of the $2^w$ configurations are identical up to cyclic permutation
and reflection symmetry.  To carry out the enumeration, we first encode
each velocity configuration as a binary sequence.  Using bit
manipulation techniques, we identify the ``irreducible'' representation
of this binary sequence, defined as the smallest equivalent integer
number obtained by performing all possible cyclic permutations of the
initial binary sequence.  This same procedure is then repeated on the
reversed initial binary sequence.  Thus to each binary sequence there is
a unique irreducible binary sequence.  By this mapping, we only need
consider the irreducible configurations and weight each by their
degeneracy in performing the average over velocity configurations.  For
example for $w=23$, 25, 27, and 29, the number of irreducible
configurations are 92,205, 337,594, 1,246863, and 4,636,390, compared,
{\it e.\ g.}, to $2^{29}=536,870,912$.  For each irreducible
configuration, we then perform the exact probability distribution
propagation.  This complete enumeration provides the exact value of
$S(t)$ for an infinite system up to $w-1$ time steps, while finite width
crossover effects gradually begin to play a role for later times.

\begin{figure}
\narrowtext
\epsfxsize=3.0in\epsfysize=3.0in
\hskip 0.05in\epsfbox{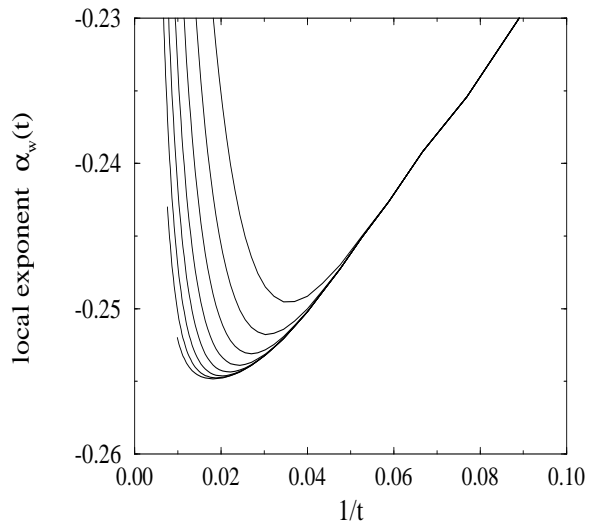}
\vskip 0.1in
\caption{The local slope of $S(t)$ versus $t$ on a double
logarithmic scale, based on averaging the exact probability distribution
over all of velocity configurations for systems of width $w=15$, 17,
$\ldots$, 29.  The data for different $w$ coincide for larger $1/t$, but
then separate at progressively smaller $1/t$ as $w$ increases
\label{fig3}}
\end{figure}

We therefore typically carried out the probability propagation for up to
$t\approx 2w$ time steps and exploited the crossover in $S(t)$ to
interpret our results.  A plot of $\ln S(t)$ versus $\ln t$ should
initially show power law behavior, indicative of the infinite system
behavior, and then cross over to a non-zero constant because of finite
width effects.  Thus a plot of the local exponent $\alpha_w(t)$ (here
defined as the slope between every other data point) versus $1/t$ should
initially provide an estimate of the exponent of $S(t)$, while the
crossover effect determines the time range over which the exact data is
relevant for the infinite system.  In Fig.~3, this local slope, is
plotted versus $1/t$ for system widths $w$ between 15 and 29.
Initially, $\alpha_w(t)$ is decreasing nearly linearly in $1/t$, but
subsequently there is the expected crossover to the asymptotic value of
zero.  In the regime where the data is relatively linear, we compute the
intercepts of successive data points at $1/t=0$ as an estimate of the
asymptotic value of the exponent (Fig.~4).  As $w$ increases, this data
exhibits: (i) non-monotonic trends in the data ({\it e.\ g.}, the
location of the minimum) which disappear only for $w\geq 25$, (ii) more
stable extrapolated values as $w$ increases, and (iii) the minimum value
of the extrapolated exponent -- which we adopt as the best estimate of
the exponent for a given value of $w$ -- is increasing systematically in
$w$ and appears to be converging to $-0.25$.  While there is slow
convergence to asymptotic behavior which gives rise to considerable
subjectivity in analysis, we believe that the trends in the data support
the hypothesis that $S(t) \sim t^{-1/4}$.

\begin{figure}
\narrowtext
\epsfxsize=3.0in\epsfysize=3.0in
\hskip 0.05in\epsfbox{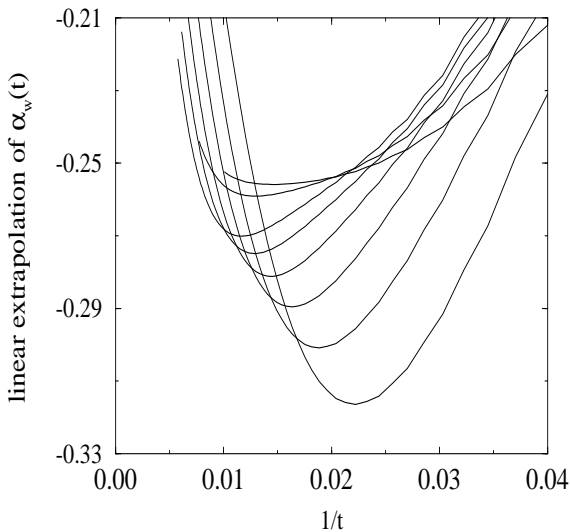}
\vskip 0.1in
\caption{Linear extrapolation of the local slope from Fig.~3 for $w=15$, 17,
$\ldots$, 29.  The minimum value is progressively increasing with $w$.
\label{fig4}}
\end{figure}

\section{THE PROBABILITY DISTRIBUTION}

In addition to investigating $S(t)$, we also examined the probability
distribution of the surviving particles.  This quantity provides an
alternative understanding for the first-passage process, as well as
useful fundamental insights about the continuum description of MdM
model.  Specifically, we study $P(x,t)\equiv \int p(x,y,t)\, dy$, the
configuration averaged longitudinal probability distribution of
particles which have not yet been absorbed by time $t$.  We expect that
this probability distribution can be written in a scaling form

\begin{equation}
\label{pxt}
P(x,t)=A f(x/\langle x\rangle),
\end{equation}
where $\langle x\rangle=\langle x(t)\rangle$ is the average longitudinal
displacement of the survivors at time $t$.  Monte Carlo simulations
clearly indicate that $\langle x\rangle\propto t^{3/4}$, as in the case
when there is no absorbing boundary present\cite{mm,bou,sr,zum}.
Because $\int_0^\infty P(x,t)\,dx= S(t)$, we can
determine the coefficient $A$ by integrating Eq.~(\ref{pxt}) over $x$ and
thereby write
\begin{equation}
P(x,t)={S(t)\over{\langle x\rangle F_0}} f(x/\langle x\rangle),
\label{apxt}
\end{equation}
where $F_0=\int_0^\infty du\,f(u)$.  Because of the absorbing boundary,
$P(x=0,t)$ must equal zero, leading to the expectation that $f(u)$ will
vanish as a power law as $u\to 0$.  Consequently, we write
\begin{equation}
P(x,t)={S(t)\over{\langle x\rangle F_0}} \left({x\over{\langle
x\rangle}}\right)^\mu g(x/\langle x\rangle),
\label{bpxt}
\end{equation}
with $g(u)\to{\rm const.}$ as $u\to 0$, and $g(u)$ vanishing faster than
any power law for $u\to\infty$.

A plot of the scaling function $f(u)$ versus $u$ is shown in Fig.~4 for
$t=255$, 1023, and 4095. There is a small but systematic variation in
the data for different times, with the small-$u$ behavior steepening and
the large-$u$ tail growing for larger time.  Nevertheless, reasonable
data collapse is obtained in which $f(u)$ qualitatively exhibits the
expected power-law and rapid cutoff asymptotic behaviors for small and
large $u$, respectively.  We attribute the small deviation from scaling
on slow convergence to the asymptotic limit (see below).  Such a
phenomenon was observed previously in the probability distribution for
an infinite system\cite{bou,sr}, and similar slow convergence effects
can be anticipated here as well.

\begin{figure}
\narrowtext
\epsfxsize=3.0in\epsfysize=3.0in
\hskip 0.05in\epsfbox{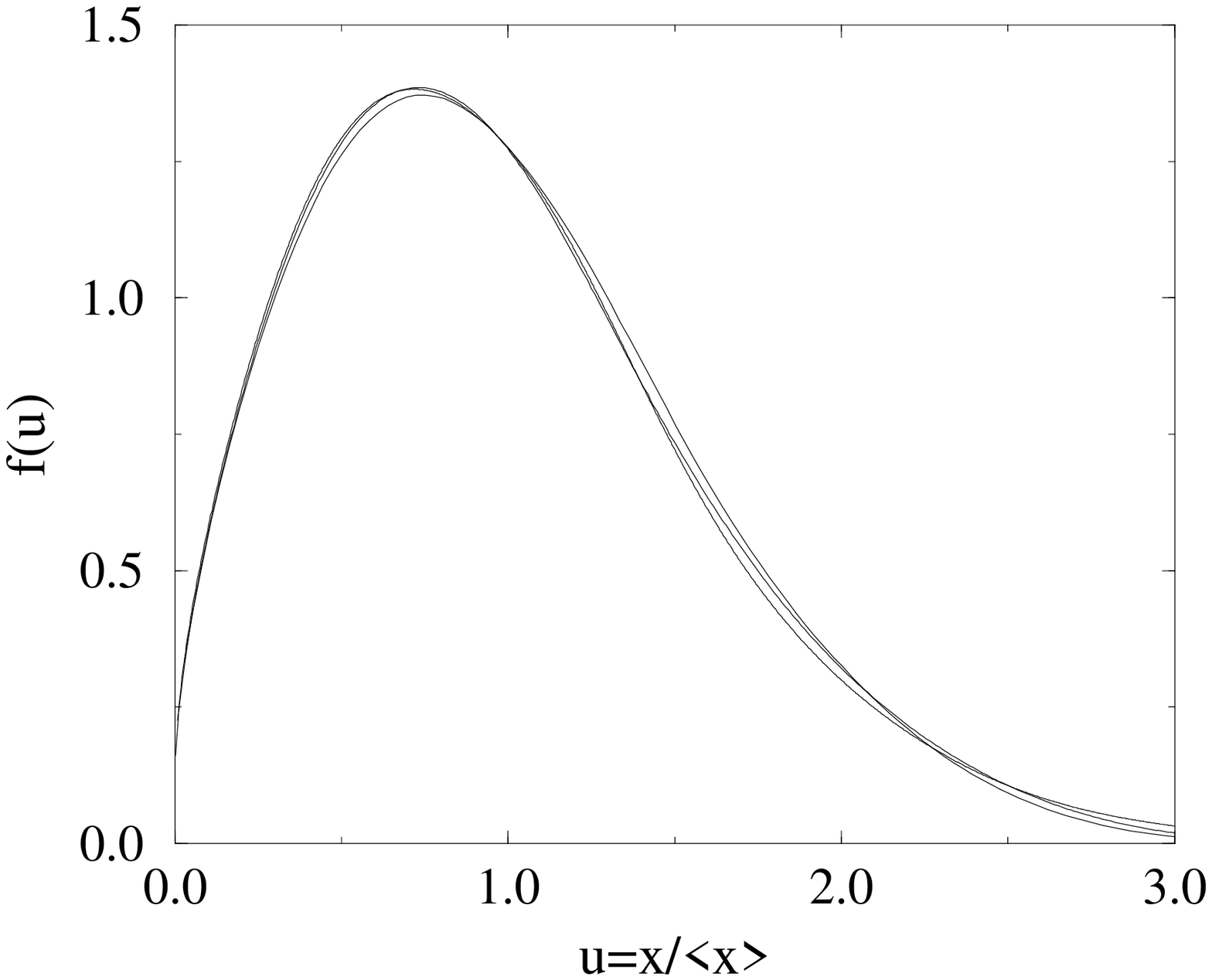}
\vskip 0.1in
\caption{The scaling function $f(u)$ versus $u$.
\label{fig5}}
\end{figure}

The exponent $\mu$ in Eq.~(\ref{bpxt}) can be obtained by demanding
consistency between the time dependence of $S(t)$ and that of the
first-passage probability.  If $S(t)\propto t^{-1/4}$, then from the
general relation\cite{rw} between $S(t)$ and the first passage
probability to the boundary, ${\cal F}(t)$, we have ${\cal
F}(t)=-dS(t)\over{dt}\propto t^{-5/4}$.  On the other hand for a
normalized initial condition, the first passage probability coincides
with the flux to $x=0$.  Now as $x\to 0$, Eq.~(\ref{bpxt}) gives
\begin{equation}
P(x,t)\propto {S(t)\over{\langle x\rangle^{1+\mu}}} x^\mu
\propto t^{-1/4-3(1+\mu)/4}\, x^\mu.
\label{flux}
\end{equation}
Since the flux is obtained by performing an appropriate spatial
derivative of this limiting probability distribution, an operation which
does not affect the temporal behavior, we conclude that $\mu=1/3$ to
recover the correct $t^{-5/4}$ time dependence for the flux.  

However, the data in Fig.~5 does not exhibit this behavior because of
finite time effects.  For small $u$, the numerical value of $g(u)$ at
$x=1$ is non-zero but decreasing with time.  Correspondingly, the value
of $u$ at this first data point is non-zero but also decreasing with
time.  This anomaly in the small-$u$ data renders a simple power law fit
inadequate.  However, such a naive fit to the data in the range $u<1/2$
gives the estimates 0.483, 0.479, 0.476, 0.470, and 0.464, respectively,
for the exponent $\mu$ in $P(x,t)$ for the 5 aforementioned time values.
Other analyses, such as computing the first derivative of $f(u)$ (which
should diverge as $u^{-2/3}$) and determining the exponent of the
small-$u$ dependence, lead to a similar quantitative conclusion.  While
the anticipated value $\mu=1/3$ is not obtained, we believe that better
agreement with theory would emerge if it were practical to extend our
Monte Carlo simulations to much longer times.

The existence of this power-law prefactor further suggests that in a
continuum description, the configuration averaged flux can be obtained via
\begin{equation}
{\cal F}(t)\,\propto 
-{\partial P(x,t)\over{\partial x^{1/3}}}\bigg|_{x=0}\;
\propto \, -x^{2/3} {\partial P(x,t)\over{\partial x}}\bigg|_{x=0}.
\label{fluxa}
\end{equation}
Such a scale-dependent diffusivity can be heuristically justified by
appealing to a Taylor diffusion description of the motion in the MdM
model\cite{taylor}.  In a time $t$, a particle typically explores
$w\approx\sqrt{Dt}$ transverse layers.  The typical bias within this
number of layers is then proportional to $w^{-1/2}$ or $(Dt)^{-1/4}$.
Thus in a time scale $t$, the typical longitudinal distance travelled by
a particle is $d\sim vt\propto t^{3/4}$.  Because these segments of
length $d$ are randomly in the $+x$ or $-x$ direction and the time
interval between segments is of order $t$, we infer an effective
longitudinal diffusion (or dispersion) coefficient $D_{||}\sim
\ell^2/t\propto t^{1/2}\propto x^{2/3}$, as written in
Eq.~(\ref{fluxa}).

Let us now pursue the consequences of this scale dependent diffusion
coefficient for the longitudinal motion.  If the longitudinal flux
$j(x,t)$ is indeed proportional to $ - x^{2/3} \partial
P(x,t)\over{\partial x}$, then substituting this into the continuity
equation
\begin{equation}
{\partial P(x,t)\over{\partial t}}+{\partial j(x,t)\over{\partial x}}=0,
\label{cont}
\end{equation}
leads to the effective equation of motion
\begin{equation}
{\partial P(x,t)\over{\partial t}}={\partial\over{\partial x}} x^{2/3} 
{\partial P(x,t)\over{\partial x}}.
\label{motion}
\end{equation}
We can easily solve this equation by applying scaling.  Assuming that
$P(x,t)\propto t^{-3/4} h(x/t^{3/4})$, we rewrite the partial
derivatives in $x$ and $t$ in terms of a derivative with respect to
$u\equiv x/t^{3/4}$, to recast the equation of motion as
\begin{equation}
-{3\over 4}(uh(u))'=(u^{2/3}h'(u))'.
\label{scaled-amotion}
\end{equation}
Here the prime denotes differentiation with respect to $u$.  One
integration immediately yields
\begin{equation}
-{3\over 4}uh(u)=(u^{2/3}h'(u)).
\label{integral}
\end{equation}
The constant of integration equals zero because $h(u)\to 0$ faster than
any power law as $u\to\infty$.  A second integration then gives
$h(u)\propto\exp(-u^{4/3})$, from which we conclude that the
longitudinal probability distribution has the form
\begin{equation}
P(x,t)\propto t^{-3/4} \exp(-(x/t^{3/4})^{4/3}).
\label{final}
\end{equation}
This functional form coincides with that obtained previously by a
different method\cite{bou,sr} in which the dominant contribution to the
large-$u$ tail of $P(x,t)$ arises from extreme ``stretched''
trajectories in unlikely velocity configurations.  Thus the observation
of the large-$u$ tail for $P(x,t)$ from the numerical data in Fig.~5 can
again be anticipated to be problematical; much more extensive simulation
would be needed.

\section{DISCUSSION}

We have investigated the time dependence of the configuration averaged
survival probability, $S(t)$, in a semi-infinite two-dimensional system
for diffusing particles which are also driven by a unidirectional random
zero-mean velocity field, $v_x(y)$.  A qualitative argument suggests
that $S(t)\propto t^{-1/4}$, a prediction which is in excellent
agreement with numerical results.  We also examined the longitudinal
spatial probability density of the surviving particles, $P(x,t)$.
Interestingly, although the numerical evidence supporting the prediction
that $ S(t)\propto t^{-1/4}$ is strong, the numerical data for $P(x,t)$
indicates slow convergence to the scaling limit and some inconsistency
with the behavior of $S(t)$ itself.  Similar anomalies in the
probability distribution occur in the unbounded
geometry\cite{bou,sr,zum}, due to the contribution of extreme velocity
configurations in the average.  It is surprising that the behavior of
$S(t)$ is apparently relatively insensitive to the contribution of such
extreme configurations.

An interesting byproduct of the spatial distribution of the surviving
particles is that the form $-\partial P(x,t)\over{\partial x^{1/3}}$ is
suggested as the appropriate expression for the particle flux.  This
leads to a scale dependent diffusion coefficient which is proportion to
$x^{2/3}$, as well as the continuum equation of motion,
Eq.~(\ref{motion}), for the longitudinal spatial probability
distribution in an unbounded geometry.  The solution to this equation of
motion is simple to obtain and reproduces the known asymptotic form of
$P(x,t)$ in the unbounded geometry\cite{bou,sr,zum}.  As an application
of this equation of motion, we find new predictions for steady-state
transport properties.  For example, for a steady input of particles at
$x=L$ and particle absorption at $x=0$, the steady solution to
Eq.~(\ref{motion}) gives a configuration-averaged density profile which
varies as $x^{1/3}$.  It will be worthwhile to test this prediction and
also the general prescription for obtaining an effective equation of
motion from the behavior of the particle flux near an absorbing
boundary.

Finally, the behavior of $S(t)$ for a semi-infinite system with a
longitudinal MdM velocity field can be easily generalized to arbitrary
spatial dimension $d$.  From classical results\cite{rw}, the number of
distinct longitudinal rows visited by a random walk in time $t$ varies
as $t^{(d-1)/2}$ for dimension $2<d<3$ ({\it i.\ e.}, transverse spatial
dimension between 1 and 2), as $t/\ln t$ for $d=3$, and as $t$ for
$d>3$.  Following closely the approach in Sec.\ II, this then leads to
\begin{equation}
\label{generald}
S(t)\sim\cases{ t^{-1/4} & $d=2$;\cr &\cr t^{-(d-1)/4} & $2<d<3$; \cr &\cr
   (\ln t/ t)^{1/2} & $d=3$;\cr &\cr t^{-1/2} & $d>3$. \cr}
\end{equation}
Thus above three dimensions, the survival probability exponent value is
not affected by the presence of a random velocity field.

\bigskip

I thank P. L. Krapivsky for many helpful discussions and a critical
reading of the manuscript.  This research was supported in part by the
NSF grant number DMR-9632059.  This financial assistance is gratefully
acknowledged.

\end{multicols} 

\end{document}